\documentclass[12pt, nature]{revtex4}
\usepackage{ifpdf}
\usepackage[english]{babel}
\usepackage{rotating}
\usepackage{graphicx}
\usepackage{color}
\usepackage{soul}
\usepackage{caption}

\expandafter\let\csname equation*\endcsname\relax
\expandafter\let\csname endequation*\endcsname\relax
\usepackage{amsmath}
\usepackage{empheq}

\begin{document}

\baselineskip24pt

\title{
Dipolar spin-waves and tunable band gap at the Dirac points in the 2D magnet ErBr$_3$
}

\author{Christian~Wessler$^{1,2*}$, 
		Bertrand~Roessli$^{1}$, 
		Karl~W.~Kr\"amer$^{3}$, 
		Uwe~Stuhr$^{1}$, 
		Andrew~Wildes$^{4}$, 
		Hans~B.~Braun$^{3}$ 
		and Michel~Kenzelmann$^{1,2}$
\\
\normalsize{$^{1}$Laboratory for Neutron Scattering and Imaging, Paul Scherrer Institut, Villigen, Switzerland}\\
\normalsize{$^{2}$Department of Physics, Universität Basel, Basel, Switzerland}\\
\normalsize{$^{3}$Department of Chemistry, Biochemistry, and Pharmacy, University of Bern, Bern, Switzerland}\\
\normalsize{$^{4}$Institut Laue-Langevin, Grenoble, France}\\
\normalsize{$^\ast$To whom correspondence should be addressed; E-mail:  christian.wessler@psi.ch}
}

\begin{abstract}
\section*{Abstract}
\noindent 
Topological magnon insulators constitute a growing field of research for their potential use as information carriers without heat dissipation.
We report an experimental and theoretical study of the magnetic ground-state and excitations in the van der Waals two-dimensional honeycomb magnet ErBr$_3$. 
We show that the magnetic properties of this compound are entirely governed by the dipolar interactions which generate 
a continuously degenerate non-collinear ground-state on the honeycomb lattice with spins confined in the plane. 
We find that the magnon dispersion exhibits Dirac-like cones when the magnetic moments in the ground-state are related by 
time-reversal and inversion symmetries associated with a Berry phase $\pi$ as in single-layer graphene. 
A magnon band gap opens when the dipoles are rotated away from this state, entailing a finite Berry curvature in the vicinity of the K and K' Dirac points.
Our results illustrate that the spin-wave dispersion of dipoles on the honeycomb lattice can be reversibly controlled 
from a magnetic phase with Dirac cones to a topological antiferromagnetic insulator with non-trivial valley Chern number.
\end{abstract}
\maketitle
\clearpage
\section*{Introduction}

Modern electronics makes use of the electronic charge to transport information. 
Because of electron scattering in metals and semiconductors, electronic devices experience resistance which results in significant energy loss. 
Due to Joule heating and dissipation miniaturisation of transistors and integrated circuits is about to reach its technological limit. 
One way of overcoming these limitations is the use of the electron spin to store and process information.
Research in the fast-developing field of spintronics mainly focuses on the development of technologies based on spin-polarised currents 
like spin valves which in the past have revolutionized the capacity of hard disk drives~\cite{Parkin2003,Hirohata2020}. 
These efforts led to the emergence of  the field of “magnonics” which promises low energy consumption.
This technology uses spin-waves ("magnons") to process and store information and exploits the advantage that magnon-based devices can be easily tuned by an external magnetic field~\cite{Kruglyak2010,Chumak2015}.
Among candidate materials for magnonic applications, magnetic van der Waals crystals are of special interest since a single layer can be exfoliated which allows to build complex heterostructures~\cite{Geim2013,Gibertini2019}. 
One example is MnPS$_3$ that is considered for magnon transport~\cite{Xing2019} and detection of antiferromagnetic states~\cite{Zhang2020}. 
A key issue in developing magnonic materials is to control the direction of propagation of the spin-waves at the surface of artificial nanostructures. 
The discovery of magnetic crystals where spin waves are robust at the surface and can propagate only along a single direction, because they are topologically protected, has prompted new ideas for future magnonic applications~\cite{Wang2018}.
Non-trivial topological states can result from spin-orbit interactions that open a gap in the magnon Dirac cone~\cite{Mook2020,Mook2014}. 
The spin-orbit coupling is at the origin of the antisymmetric Dzyaloshinskii-Moriya (DM) interaction in magnetic crystals~\cite{Moriya1960} which induces the magnon-Hall effect observed in ferromagnetic insulators~\cite{Onose2010,Zhang2013}.
Recently, it was proposed that dipolar interactions have a similar effect as the DM interaction~\cite{Shindou2013,Shen2020}
 which paves the way for the realisation of topological artificial nano-structures~\cite{Liu2020,Liu2020-2}.

\noindent 
Dipolar interactions are always present in magnetic solids.
They are direction dependent and long-range in contrast to exchange interactions.
This property induces frustration in the lattice as in general it is not possible to satisfy all interactions simultaneously. 
The problem of determining whether a lattice of ordered magnetic dipoles can exist was addressed by Luttinger and Tisza~\cite{Luttinger1946}. 
They demonstrated that the face-centred and body-centred cubic lattices display ferromagnetism at low temperature while for the simple cubic lattice the classical ground-state is a collinear antiferromagnet.
The Mermin-Wagner theorem~\cite{Mermin1966} does not apply in the case of long-range potentials and the absence of order in two-dimensional (2D) magnets with dipolar interactions does not generally apply.
For a triangular lattice with solely dipolar interactions~\cite{Malozovsky1990}, indeed long range order is predicted. 
For both the square and honeycomb dipolar lattices, the classical ground-state exhibits a continuous degeneracy~\cite{Zimmerman1988,Rozenbaum1995}.
Finally, in the geometrically frustrated Kagome lattice no ground-state satisfies the condition that the spin length is constant on all lattice points and long-range ferrimagnetic order is predicted with numerical simulations~\cite{Maksymenko2015}.

\noindent In this work, we report an analysis of the magnetic ground-state and excitations in the van der Waals magnet $\rm ErBr_3$ which was shown to undergo a magnetic phase transition to a continuously degenerate non-collinear 2D-order below $T$ = 280~mK~\cite{Kraemer1999}. 
We show that the magnetic ground-state in this crystal is explained by dipolar interactions between the Er$^{3+}$ moments. 
Calculation of the dispersion of the magnetic excitations using the mean-field random-phase approximation (MF-RPA), that includes dipolar interactions and crystalline electric-field (CEF) anisotropy, describes the inelastic neutron-scattering (INS) data.
Using our simulations of the magnetic excitations for the dipolar honeycomb lattice, 
we show the presence of a Dirac cone at the K-points of the Brillouin zone when the configuration of dipoles possess inversion (I) and time-reversal (T) symmetry.
A gap opens at the Dirac points between the acoustic and lowest optical branch as the degeneracy angle between spins in the two sublattices of the honeycomb structure is varied and the IT symmetry of the magnetic structure is lifted.

\noindent 
\section*{Results}
\subsection*{\textnormal{ Crystal and Magnetic Structure of E\lowercase{r}B\lowercase{r}$_3$} }
\begin{figure}[ht]
\begin{center}
\includegraphics*[width=\textwidth]{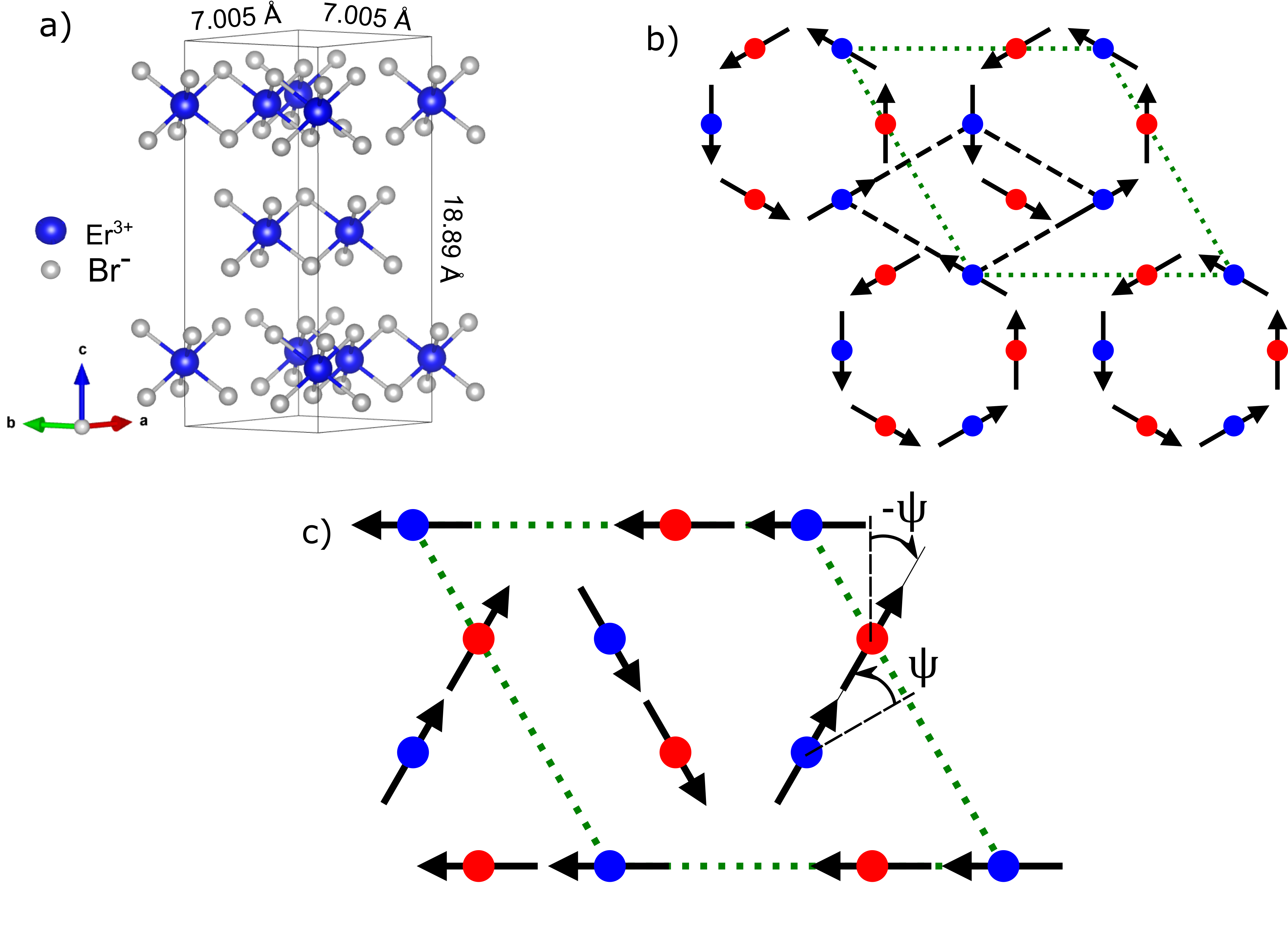}
\caption{
		a) Crystal structure of ErBr$_3$. Br$^-$ ions octahedrally coordinate Er$^{3+}$ ions which form honeycomb layers stacked along
		 the c-axis.
		b) Honeycomb layer of Er$^{3+}$ ions at $z$ = 1/6 with ordered magnetic moments of 4.7 $\mu_B$ (black arrows) for a 
		spin configuration given by angle $\mathit\Psi$ = 0. 
		The two trigonal sublattices of the honeycomb are shown as blue and red spheres. 
		The crystallographic and magnetic unit cells are shown as black and green lines, respectively.
		c) The classical magnetic ground-state is continuously degenerate
			 and the magnetic moments can be rotated in opposite direction by varying the angle $\mathit\Psi$. 
			 In panel c), $\mathit{\Psi} = \pi/6$. 
}
\label{ErBr3_Fig1}
\end{center}
\end{figure}

\noindent 
ErBr$_3$ crystallizes in the BiI$_3$ structure~\cite{Braekken1930} with the rhombohedral space group R-3 (no. 148) and lattice parameters $a$ = $b$ = 7.005 $~\rm\AA$ and $c$ = 18.89  $~\rm\AA$ at $T$ = 1.5 K~\cite{Kraemer1999}, as shown in Fig.~\ref{ErBr3_Fig1}a. 
In this layer-type structure the Br$^-$ ions adopt a hexagonal densest packing (AB) and Er$^{3+}$ ions occupy in an ordered way 2/3 of the octahedral voids between every other layer (2/3c). 
The layer stacking along the c-axis results in a sequence A2/3cBA2/3cBA2/3cB with three Er$^{3+}$ honeycomb layers per unit cell. 
The Er$^{3+}$ ions are located on site (6c) at (0,0,$\pm$z), (0,0,$\pm$z)+(2/3,1/3,1/3), and (0,0,$\pm$z)+(1/3,2/3,2/3). 
In this crystal z = 1/6 and the rare-earth ions form a perfect honeycomb lattice. 
The magnetic structure of ErBr$_3$ was determined by powder neutron diffraction~\cite{Kraemer1999}. 
Magnetic order sets in below $T$ = 280~mK and down to $T$ = 50~mK only two-dimensional (2D) order is observed. 
The magnetic ordering vector ${\textbf{k}}_{0}$ = (1/3,1/3,0) triples the volume of the magnetic unit cell to $\sqrt{3}a \times \sqrt{3}a \times c$ with axes rotated by 30$^{\circ}$ from the crystallographic cell, see Fig.~\ref{ErBr3_Fig1}b. 
The ordered magnetic moments are located in the honeycomb plane and amount to $\mu$ = 4.7 $\mu_B$ at $T$ = 50~mK. 
The honeycomb lattice is composed of two trigonal sublattices. 
The Er$^{3+}$ ions on each sublattice exhibit a 120$^{\circ}$ antiferromagnetic order with opposite chirality, see Fig.~\ref{ErBr3_Fig1}b. 
The magnetic ground state has an infinite degeneracy with respect to in-plane rotation of the ordered moments by an arbitrary angle $\mathit\Psi$ with opposite signs for the two sublattices~\cite{Kraemer1999}, see Fig.~\ref{ErBr3_Fig1}c.

\noindent 
In the mean-field approximation the magnetic ground-state is determined by the eigenvector of the largest eigenvalue of the Fourier transform of the exchange interaction tensor~\cite{Bertaut1963}. 
The Hamiltonian with purely dipolar interactions is written
\begin{equation}
H = -\frac{(g\mu_B)^2\mu_0}{8\pi}\sum_{i,j}{ \textbf{J}} _i\cdot{\overline{\overline D}}(ij)\cdot{ \textbf{J}}_j
\end{equation}
and
\begin{equation}
D_{\alpha,\beta}(ij)=\frac{3({ \textbf{ R}}_{i}-{ \textbf{ R}}_{j})_{\alpha}({ \textbf{ R}}_{i}-{ \textbf{ R}}_{j})_{\beta}}{|{ \textbf{ R}}_{i}-{ \textbf{ R}}_{j}|^{5}}
-\frac{ \delta_{\alpha,\beta} }{|{ \textbf{ R}}_{i}-{ \textbf{ R}}_{j}|^{3}}.
\end{equation}
$\alpha, \beta = {a,b,c}$ are the crystal directions in Cartesian coordinates and ${ \textbf{ R}}_i$ is the position vector of the $i$ magnetic ion. 
The dispersion of the eigenvalues of the Fourier transform of the dipolar tensor along $\rm { \pmb \kappa }=(q,q,0)$ for the $\rm ErBr_3$ crystal with six Er-ions in the chemical cell is shown in Supplementary Figure 1 of the Supplementary Note 1 in the Supplementary information (SI). 
The largest eigenvalue has a maximum at ${ \textbf{ k}} _0= (1/3,1/3,0)$ in agreement with the propagation vector of the magnetic structure measured in ErBr$_3$. 
This particular non-collinear magnetic ordering cannot be reproduced by frustrated Heisenberg interactions \cite{Rastelli1999}.
The temperature of the magnetic phase transition calculated with mean-field theory~\cite{Enjalran2004} $ \rm \textit{T}_{MF} = max\{\lambda_{{ \textbf{ k}}_0}/k_B\}/3 \sim 0.3~K$ is close to the ordering temperature of the Er moments measured by neutron diffraction.   
Mean-field theory predicts the same magnetic ground-state that is observed in ErBr$_3$ and agrees with the spin order obtained by Rozenbaum~\cite{Rozenbaum1995} for magnetic dipoles on the bipartite honeycomb lattice.
The magnetic ground-state is continuously degenerate and the equivalent spin structures are defined in terms of an arbitrary angle $\mathit\Psi$, as shown in Fig.~\ref{ErBr3_Fig1}c. 
In the SI, Supplementary Figure 2 shows the dependence of the ground-state energy upon the lattice parameter $c$ of the crystal structure of ErBr$_3$. 
The dipolar energy is constant for a lattice parameter \textit{c} $\rm > 10 ~\AA$ which indicates that interlayer interactions become negligibly small above that value. 
With a lattice parameter \textit{c} = 18.89 ~$\rm \AA$ and an interlayer distance of \textit{c}/3 = 6.3$~\rm\AA$ the 2D limit is reached in $\rm ErBr_3$.
\subsection*{\textnormal{Spin Waves} }
To calculate the dispersion of the magnetic excitations we used the MF-RPA method which treats the single-ion interactions (crystal-field and mean-field Hamiltonians) exactly while the interactions between the magnetic ions are considered in the random-phase approximation~\cite{Jensen1991}. 
The inelastic neutron scattering measurements of the crystal-field splitting of ErBr$_3$ are presented in the Supplementary Note 2 and the analysis of the data explains the temperature dependence of the single-ion susceptibility shown in Supplementary Figure 5.
In $\rm ErBr_3$, the first crystal-field level is located at 1.5 meV (see Supplementary Figure 5 in the SI), 
much higher than the energy scale of the dipolar interactions;
the spin system can be approximated by a $S$ = 1/2 doublet with effective anisotropic $g$-factors as detailed in Supplementary Note 3.
For ErBr$_3$, we obtained a large easy-plane anisotropy with $g_x = g_y = 9.4$ and $g_z = 0.1$.

\noindent 
In MF-RPA the spin-wave excitations appear as poles in the dynamical susceptibility tensor $\overline{\overline \chi}( {\textbf{q}}, E) $,
\begin{align}
\begin{aligned}
\overline{\overline \chi}({\textbf{q}},E) = [\overline{\overline 1} -\overline{\overline \chi}_0(E)\overline{\overline D}({\textbf{q}})]^{-1}\overline{\overline \chi}_0(E),
\end{aligned}
\end{align}
where $\overline{\overline \chi}_0(E)$ is the single-ion susceptibility and $\overline{\overline D}( {\textbf{q}})$ the Fourier transform of the dipolar interactions.
%
Because the magnetic moments of $\rm ErBr_3$ in the 2D phase form a non-collinear structure, it is convenient to introduce local coordinates for each spin 
in the magnetic cell with the quantization axis pointing along the spin direction, 
so that after transformation the single-ion susceptibility is the same for every magnetic site in the unit cell~\cite{Jensen1991}.
The inelastic neutron cross-section $S({\pmb{ \kappa}}, E)$ is then proportional to the imaginary part of the dynamical susceptibility 

\begin{align}
S({\pmb \kappa},E) = &\frac{1}{\pi}\frac{1}{1-\exp(-E/k_BT)} \frac{1}{N}\sum_{\alpha,\beta}\sum_{u,v}(\delta_{\alpha,\beta}-\frac{\kappa_\alpha\kappa_\beta}{| {\pmb\kappa}|^2}) \Im\chi^{\alpha,\beta}_{u,v}( {\pmb\kappa}, E)
\label{ncs}
\end{align}
with ${ \pmb{\kappa}} ={ \pmb{\tau}} + {\textbf{q}}$~ the scattering vector and ${\pmb{ \tau}}$ a reciprocal lattice vector; $u,v$ number the $N$ Er-ions 
in the magnetic cell and $\alpha,\beta=a,b,c$. 
Eq.~\ref{ncs} is given in the crystal frame and hence, the $\Im\overline{\overline{\chi}}_{u,v}( \kappa, E)$ matrices 
are evaluated in the rotated local coordinate system.

\noindent 
For the spin-wave calculations, only interactions within a single honeycomb layer were considered as the dipolar interactions between the planes are negligibly small. 
The spin wave dispersion is shown in Supplementary Figure 7 in the SI. For scattering vectors along $(0,q,0)$ an acoustic mode and three optic branches are well observable, while the other two spin-wave branches have a considerably lower intensity. 
Along this direction in reciprocal space $S({\pmb \kappa},E)$ does not change appreciably when $\mathit\Psi$ is varied 
for the optical branches above $E \simeq$ 0.1 meV
and no crossing of the acoustic mode with an optical branch occurs.

\begin{figure}[h]
\begin{center}
\includegraphics*[width=\textwidth]{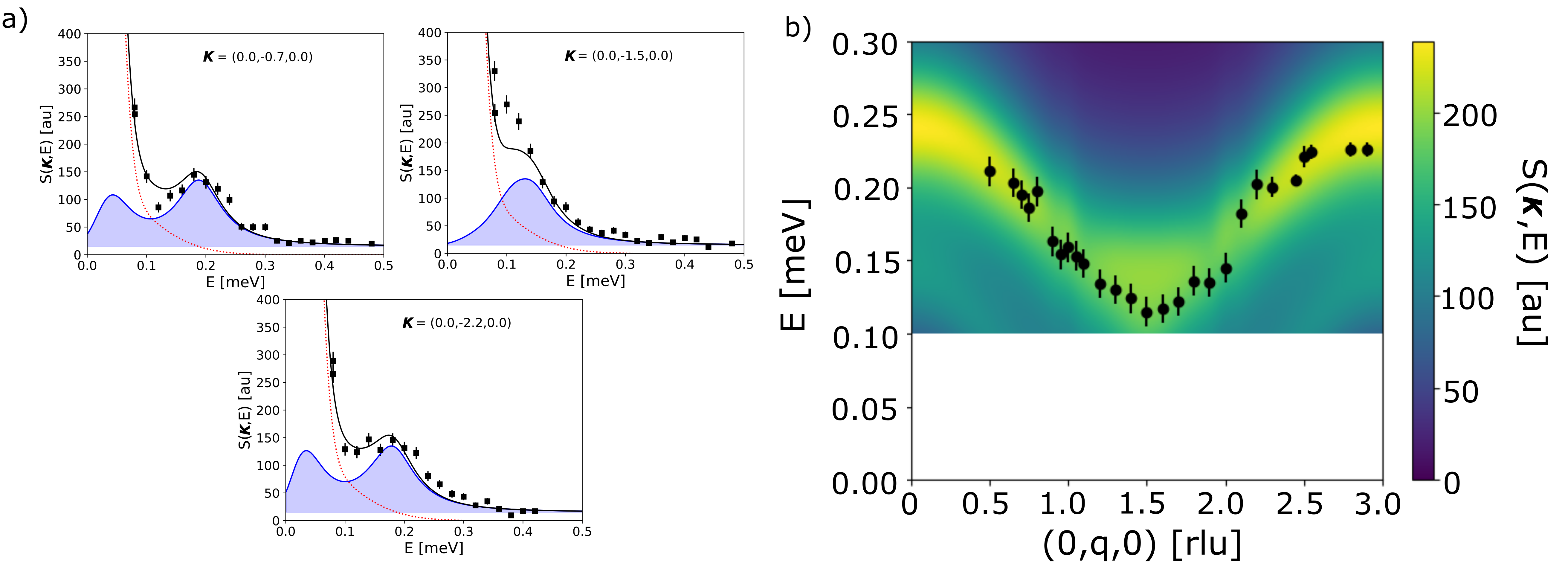}
\caption{a) Representative inelastic neutron scattering measurements of the spin-wave excitations in $\rm ErBr_3$ compared with the result of 
		mean field-random phase approximation (blue lines).
		The red line is a fit of the incoherent scattering measured at momentum transfer ${\pmb \kappa}$ = (2/3,-4/3,0) and at $T$ = 6~K, 
		which is well above the temperature of the phase transition $T$ = 280 mK and in the paramagnetic phase 
		where there are no propagating coherent spin waves.
		The sum of both contributions is shown by the black line.
		Error bars are standard deviations.
	     b) False color plot of the inelastic neutron cross-section $S({\pmb \kappa},E)$ along ${\pmb \kappa} =(0,q,0)$.
		     In the calculations, we used a linewidth $\gamma$ = 0.07 meV in the single-ion susceptibility to approximate the energy resolution of the 
		     spectrometer (see Supplementary Note 3).
		The points represent the spin-wave dispersion in ErBr$_3$ obtained from a fit with a single Gaussian. 
		Magnetic excitations below the energy $E$ = 0.1 meV were not accessible in the experiment.
		The error bars are a result from the Gaussian fit.
		}
\label{spinwavescomp}
\label{ErBr3_Fig2}
\end{center}
\end{figure}

\noindent The inelastic neutron scattering (INS) measurements in $\rm ErBr_3$ were performed with an energy resolution of about $\rm 70 ~\mu eV$  and the spin-waves that have an energy $ E <$ 0.1~meV are hidden by the incoherent scattering at the elastic position. 
Fig.~\ref{spinwavescomp}a shows representative INS measurements along $(0,q,0)$ at $T$ = 80 mK together with the result of MF-RPA.
Additional INS measurements are shown in Supplementary Figure 6 together with a plot of the magnetic Brillouin zone where the scan directions are indicated. 
Only an overall scale factor is used to adjust the INS intensity with the MF-RPA calculations. 
The peaks in the INS spectrum at $\sim$ 0.2 meV correspond to the dispersion of the highest visible magnon branch. 
Both the dispersion of the spin waves and the intensity of the INS are well reproduced.
In Fig.~\ref{spinwavescomp}b, we compare the dispersion of the spin-waves in $\rm ErBr_3$ with the MF-RPA calculations. 
It can be observed that the calculated spectrum of $S({\pmb \kappa}, E)$ is in agreement with the INS data.
\section*{Discussion}

The electronic band structure of materials with a honeycomb structure, like graphene, exhibits Dirac points at the corners of the Brillouin zone 
(the K-points in Supplementary Figure 6c in the SI) where the electronic bands cross. 
Close to the K-point the electronic bands have a linear dispersion that corresponds to the dispersion of a relativistic and massless particle~\cite{Wehling2014}. 
It was shown that the spin-waves of a Heisenberg ferromagnet with a honeycomb structure have Dirac cones~\cite{Franson2016,Boyko2018} around the K point opening the possibility of producing magnetic Dirac materials. 
In agreement with these predictions topological magnons have been observed as Dirac massless bosons in the ferromagnet CrBr$_3$~\cite{Pershoguba2018}. 
The acoustic and optical spin-wave branches cross at the K point of the Brillouin zone boundary and have a linear dispersion analogous to the electronic band structure of graphene. 
Magnon Dirac cones are not only found in ferromagnetic insulators and topological magnons protected by symmetry have been observed in the 3D antiferromagnets Cu$_3$TeO$_6$~\cite{Li2017,Bao2018,Yao2018} and CoTiO$_3$~\cite{Yuan2020}. 
In contrast, a magnon band gap is observed at the zone boundary in CrI$_3$.
The origin of the gap is still debated and has been associated either with the Dzyaloshinskii-Moriya interaction between next-nearest
neighbours on the honeycomb lattice ~\cite{Chen2018} or magnon-phonon coupling \cite{Delugas2021}.
In magnetic insulators, the DM interaction plays the role of the spin-orbit interaction present in graphene and leads to avoided band crossing of magnons.
According to the bulk-edge correspondence~\cite{Hasan2010},  magnons, in the gapped phase, can propagate uni-directionally along the surface or edges which opens the possibility of producing magnetic counterparts of topological insulators.

\begin{figure}[h]
\begin{center}
\includegraphics*[width=\textwidth]{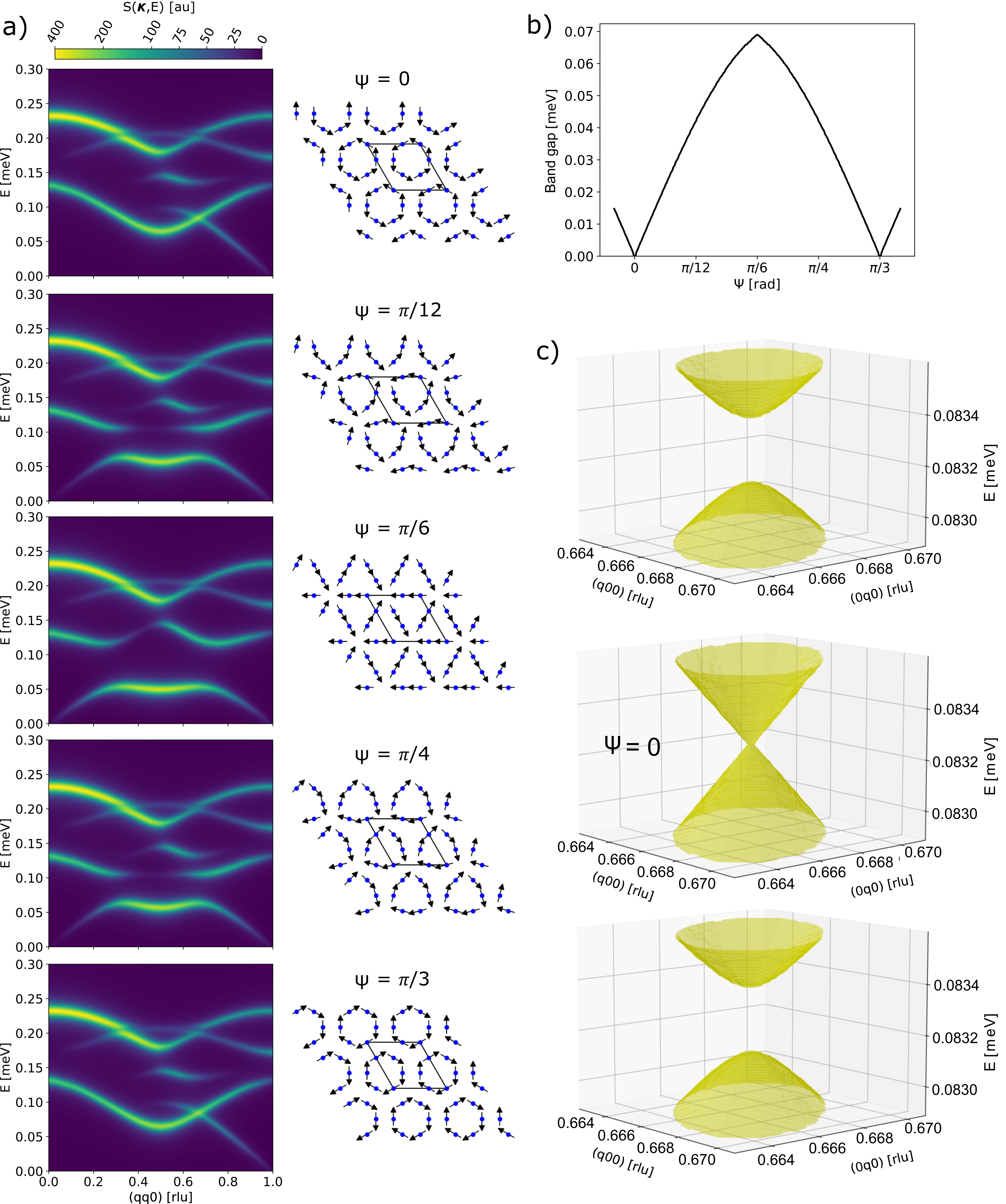}
\caption{ a) Simulation of the inelastic neutron cross-section $S({\pmb \kappa}, E)$ along the momentum transfer 
		${\pmb \kappa}=(q,q,0)$ for selected $\mathit\Psi$.  
   		 Note the opening of a gap at $ {\pmb\kappa}=(2/3,2/3,0)$ between the acoustic and the lowest optical mode at the energy
   		 $ E \approx$ 0.083~meV when the spin configuration given by angle $\mathit\Psi$ is varied.
   		 The colour bar represents the calculated intensity of the inelastic neutron cross-section.
   		  b) Dependence of the magnon band gap as a function $\mathit\Psi$ at $ {\pmb\kappa}=(2/3,2/3,0)$.	 
   		  c) Calculation of the spin-wave dispersion close to the Dirac point for $\mathit{\Psi} = -5.6 \times 10^{-4}$  $\pi$, 0, and +5.6 $\times 10^{-4}$ $\pi$, 				respectively.  
   		 }
\label{ErBr3_Fig3}
\end{center}
\end{figure}

Fig.~\ref{ErBr3_Fig3}a shows the dispersion of the spin waves for ErBr$_3$ along the (q, q, 0) direction. 
As the magnetic ground-state is degenerate, we have performed the calculations for different values of $\mathit\Psi$ for the magnetic moments of
the Er-sublattices.
The dispersion of the spin waves for the dipolar honeycomb lattice shows a Dirac cone at the K-point when the spins in ErBr$_3$ form a periodic vortex structure with $\mathit\Psi$ = 0 (Fig.~\ref{ErBr3_Fig3}a top). 
When $\mathit\Psi$ is varied an anti-crossing of the lowest optical branch with the acoustic branch appears at the K-point of the magnetic Brillouin zone. 
The size of the energy gap between the two branches depends on $\mathit\Psi$ as shown in Fig.~\ref{ErBr3_Fig3}b and reaches a maximum for $\mathit{\Psi} = \rm{\pi}/6$. 
The splitting between the acoustic and optic branch decreases again for values of $\mathit\Psi$ larger than $\pi/6$; finally the two branches form a Dirac cone again when the arrangement of the Er-moments corresponds to a periodic vortex structure with $\mathit\Psi$ = $\pi/3$ (Fig.~\ref{ErBr3_Fig3}a bottom).
Furthermore, the variation of $\mathit\Psi$ from 0 to 2$\pi$ produces six distinct spin structures on the honeycomb net corresponding to three vortex positions with opposite chiralities  where the Dirac cones are present in the spin-wave spectrum at the K and K' points.

\begin{figure}[ht]
\begin{center}
\includegraphics*[width=\textwidth*2/3]{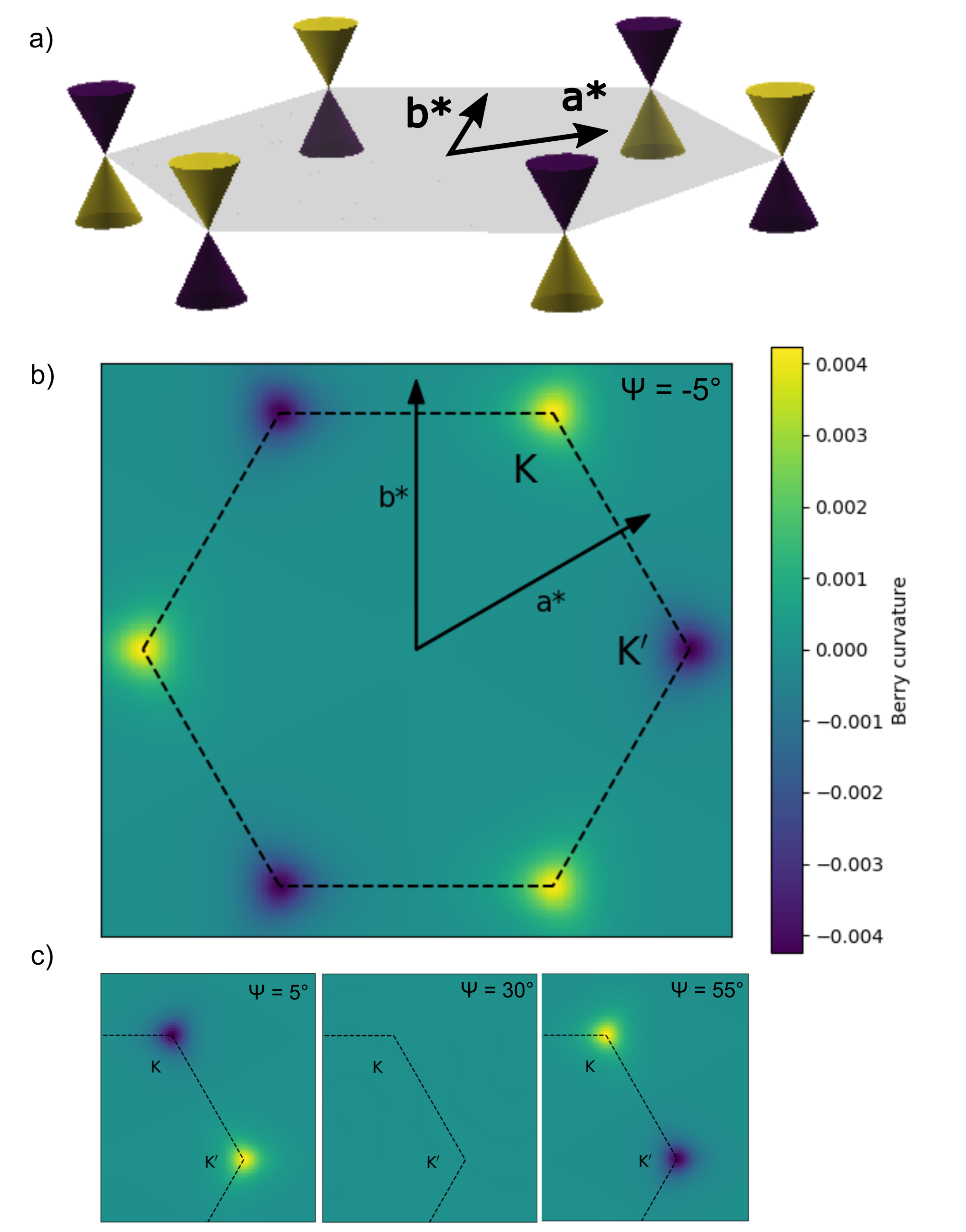}
\caption{ 
		a) The Berry phase associated with the Dirac cone ($\mathit{\Psi} = 0^{\circ}$) is either +$\pi$ (represented in yellow) or -$\pi$ (blue).
		The Berry curvature $\Omega( {\pmb k})$ in ErBr$_3$ calculated for the acoustic magnon branch:
			b) $\mathit{\Psi} = -5^{\circ}$, 
			c) $\mathit{\Psi} = 5^{\circ}$, $\mathit{\Psi} = 30^{\circ}$ and $\mathit{\Psi} = 55^{\circ}$.
		$\mathit{\Psi}$ is the angle representing the spin configuration.
		The Brillouin zone boundary is outlined with dashed lines.
		The colour bar shows the magnitude of the Berry curvature.
   		 }
\label{ErBr3_Fig4}
\end{center}
\end{figure}

For $\mathit{\Psi} = n \times \pi/3$ ($n$ an integer), 
the acoustic and the lowest optical spin-wave branch cross at the K-point with a linear slope and a Dirac cone is found
at an energy of $E \approx$ 0.083 meV, as shown in Fig.~\ref{ErBr3_Fig3}c. 
We calculated a Berry phase $\gamma=\pm\pi$ (see Fig.~\ref{ErBr3_Fig4}a and Supplementary Note 4) which reflects the degeneracy of the two magnon branches at the Dirac points and shows that the topology of the magnon dispersion in ErBr$_3$ mimics 
the topology of the electronic band structure in graphene.
Our results demonstrate that Dirac cones are not only present in honeycomb magnets with Heisenberg interactions, but also in van der Waals crystals like ErBr$_3$, where long-range dipolar forces are the primary magnetic interaction.

As in Cu$_3$TeO$_6$, Dirac points are present in the spectrum of magnetic excitations due to the invariance of the magnetic structure under combined inversion and time-reversal symmetry operations~\cite{Li2017,Bao2018,Yao2018}. 
When $\mathit{\Psi} \neq n \times \pi/3$ (n an integer), the symmetry of the spin structure is broken which generates a finite Berry curvature as shown in Fig.~\ref{ErBr3_Fig4}b. 
In Fig.~\ref{ErBr3_Fig4}c and Supplementary Figure 8, we show the evolution of the Berry curvature calculated for the magnon acoustic branch which is well separated from the other branches.
The Berry curvatures at the K and K' points have the same magnitude but opposite signs as it is the case in graphene~\cite{Xiao2007}. 
For degeneracy angles close to $\mathit{\Psi} = 0$, the Berry curvature $\Omega ( {\pmb k})$ is peaked around the K and K' points 
and diverges when $\mathit{\Psi} \rightarrow 0$. 
With increasing values of $\mathit\Psi$ we observe that $\Omega  ( {\pmb k})$ continuously broadens and weakens but remains centered around the K and K' points. 
The Berry curvature finally disappears at $\mathit{\Psi} = \pi/6$ when the gap between the acoustic and optical branch has its maximum value. 
Upon increasing the angle $\mathit\Psi$ further, the Berry curvature reappears close to the K and K' points albeit with opposite signs. 
The Berry curvature of the dipolar honeycomb magnet hence depends on the size of the spin gap and can be manipulated by varying the angle $\mathit\Psi$. 
The Chern number $C$ is equal to 0 when the Berry curvature is integrated over the entire Brillouin zone. 
Although $C=0$ indicates a topologically trivial magnon insulator phase, when the integration of $\Omega({ \pmb k})$ is restricted around a single Dirac point we numerically obtain for $\mathit{\Psi} = 1.5^{\circ}$ $C_V \approx \pm 1/2$ at the K and K' points which is a prerequisite for the quantum valley Hall effect~\cite{Yao2009, Kai2018}. 
A non-trivial Berry curvature is also at the origin of the magnon valley transports that were recently reported in Ref.~\cite{Ghader2020,Zhai2020}.
For the dipolar honeycomb lattice, a topological phase transition occurs at $\mathit\Psi$ = 0 where the spin gap closes at the Dirac points and \textit{C}$_V$ changes sign.
As the Berry curvature is anti-symmetric with respect to the transition point, a 
topological magnon can propagate along the interface of two spin-domains located on opposite sides of the phase boundary~\cite{Mook 2015}.
It is worth restating that these considerations hold for purely dipolar interactions.
If exchange interactions are present, long-range ordering will occur with the moments normal to the plane~\cite{Pich 1995}.

In conclusion, we have shown that 
ErBr$_3$ serves as a model crystal for the dipolar antiferromagnet on the honeycomb lattice. 
The classical magnetic ground-state forms a degenerate non-collinear vortex structure and we have found that the lowest magnon bands evolve from Dirac magnon excitations to a topological phase with non-trivial Berry curvature by varying the phase $\mathit\Psi$ that defines the spin configuration. 
The application of an external magnetic field lifts the degeneracy of the ground-state 
and can be used to select a particular spin configuration by varying the direction of the magnetic field~\cite{Zimmerman1988}.
As van der Waals magnets can  be cleaved this could be relevant for future investigations of magnon currents in honeycomb lattices. 
Our results suggest that
if artificial dipolar honeycomb lattices can be built, it would have the potential to be extremely useful for magnonic devices since 
the energy at which the magnon band crossing occurs
can be adapted by varying the size of the magnetic moment or the lattice spacing which allows to engineer the magnon band structure~\cite{Streubel2018,Lenk2011}.
\clearpage
\section*{Methods}
\subsection*{Single crystal of ErBr$_3$ and characterization}
\noindent
The single crystal of ErBr$_3$ was synthesized according to the method described previously in Ref.~\cite{Kraemer1999}. 
The magnetic susceptibility was measured with a MPMS-5XL SQUID system (Quantum Design).

\subsection*{Neutron experimental setup}
\noindent
The CEF splitting was determined with the thermal-neutron spectrometer Eiger at the SINQ spallation source. 
The spectrometer was operated in the constant final energy mode with $\rm \textit{k}_f = 2.662~\AA^{-1}$. To maximize the intensity the
monochromator was double focused and the analyzer was horizontally focused. 
With this configuration the energy resolution is about $\Delta E$ = 0.8~$\rm meV$ at the elastic position. Contamination by higher-order neutron wavelengths was eliminated by a PG002-filter installed in the scattered beam. 
The dispersion of the magnetic excitations was measured with the cold-neutron three-axis spectrometer IN14 at the Institut Laue-Langevin. 
The spectrometer was operated in a similar configuration as for the previous measurements albeit with $\rm \textit{k}_f=1.15~\AA^{-1}$ which resulted in an improved energy resolution of 70~$\rm \mu eV$. 
A cold Be-filter, that scatters off neutrons with wavevectors longer than $\rm \textit{k}_f=1.55~\AA^{-1}$, was used to reduce the background and avoid spurious scattering. 
For these measurements a single crystal of approximately 0.5 cm$^3$ was mounted inside a dilution refrigerator and cooled down to the base temperature of $T$ = 80~mK which is well below the ordering temperature of the Er$^{3+}$ moments at $T$ = 280~mK. 
\clearpage
\section*{\textsf{Data availability: }} 
All data needed to evaluate the conclusions in the paper are present in the paper and/or the Supplementary Information. 
Additional data related to this paper may be requested from the authors.
\clearpage
\section*{\textsf{Acknowledgments}}
\noindent We would like to thank C. Mudry for useful discussions.
The financial support by the Swiss National Science Foundation under grant no. SNF 200020$\_$172659
is gratefully acknowledged.
\section*{\textsf{Author contributions:}}
\noindent The single crystal was grown by K.W.K. 
The inelastic neutron scattering experiments were performed by H.B.B., K.W.K., B.R., U.S., A.W., and C.W. 
The susceptibility measurements were done by K.W.K. 
Modeling and simulations were performed by B.R. and C.W. in discussion with H.B.B., K.W.K., and M.K. 
All authors contributed to the writing of the manuscript.
\section*{\textsf{Competing interests:}}
\noindent The authors declare no competing interests.	
\clearpage

%

\clearpage

%

\clearpage
\section*{References}
\begin{thebibliography}{99}
\bibitem{Parkin2003} Parkin, S., Jiang, X., Kaiser, C., Panchula, A. \& Samant, M.
				Magnetically engineered spintronic sensors and memory.
				\textit{Proc. IEEE} \textbf{91}, 661 (2003).
\bibitem{Hirohata2020} Hirohata, A. \textit{et al.} 
					Review on spintronics: Principles and device applications. 
					\textit{JMMM} \textbf{509}, 166711 (2020).
\bibitem{Kruglyak2010} Kruglyak, V.V., Demokritov, S.O. \& Grundler, D.
				Magnonics.
				\textit{J. Phys. D: Appl. Phys.} \textbf{43}, 264001 (2010).
\bibitem{Chumak2015} Chumak, A.V., Vasyuchka, V.I., Serga, A.A. \& Hillebrands, B.
				Magnon spintronics. 
				\textit{Nat. Phys.} \textbf{11}, 453 (2015).				
\bibitem{Geim2013} Geim, A.K. \& Grigorieva, I.V. 
				Van der Waals heterostructures.
				\textit{Nature} \textbf{499}, 419 (2013).
\bibitem{Gibertini2019} Gibertini, M., Koperski, M., Morpurgo, A.F. \& Novoselov, K.S.
				Magnetic 2D materials and heterostructures.
				\textit{Nat. Nanotechnol.} \textbf{14}, 408 (2019).
\bibitem{Xing2019} Xing, W. \textit{et al}. 
				Magnon transport in quasi two-dimensional van der Waals antiferromagnets.
				\textit{Phys. Rev. X} \textbf{9}, 011026 (2019).
\bibitem{Zhang2020}  Zhang, Y. \textit{et al}. 
				MnPS$_3$ spin-flop transition-induced anomalous Hall effect in graphite flake via van der Waals proximity coupling. 
				\textit{Nanoscale} \textbf{12}, 23266 (2020).
\bibitem{Wang2018} Wang, X.S., Zhang, H.W. \& Wang, X.R. 
				Topological magnonics: A paradigm for spin-wave manipulation and device design,  
				\textit{Phys. Rev. Applied} \textbf{9}, 024029 (2018).
\bibitem{Mook2020}  Mook, A., Plekhanov, K., Klinovaja, J. \& Loss, D. 
				Interaction-stabilized topological magnon insulator in ferromagnets.
				\textit{Phys. Rev. X} \textbf{11}, 021061 (2021).
\bibitem{Mook2014} Mook, A., Henk, J. \& Mertig, I. 
				Edge states in topological magnon insulators. 
				\textit{Phys. Rev. B} \textbf{90}, 024412 (2014).
\bibitem{Moriya1960} Moriya, T.
				Anisotropic superexchange interaction and weak ferromagnetism.
				\textit{Phys. Rev.} \textbf{120}, 91 (1960).
\bibitem{Onose2010} Onose, Y. \textit{et al.} 
				Observation of the magnon Hall effect.
				\textit{Science} \textbf{329}, 297 (2010).
\bibitem{Zhang2013} Zhang, L., Ren, J., Wang, J.S. \& Li, B.
				Topological magnon insulator in insulating ferromagnet.
				\textit{Phys. Rev. B} \textbf{87}, 144101 (2013).
\bibitem{Shindou2013} Shindou, R., Ohe, J, Matsumoto, R., Murakami, S. \& Saitoh, E.
				Chiral spin-wave edge modes in dipolar magnetic thin films. 
				\textit{Phys. Rev. B} \textbf{87}, 174402 (2013).
\bibitem{Shen2020} Shen, K.
				Magnon spin relaxation and spin Hall effect due to the dipolar interaction in antiferromagnetic insulators.
				\textit{Phys. Rev. Lett.} \textbf{124}, 077201 (2020).
\bibitem{Liu2020}  Liu, J., Wang, L. \& Shen, K.
				Spin-orbit coupling and linear crossings of dipolar magnons in van der Waals antiferromagnets.
				\textit{Phys. Rev. B} \textbf{102}, 144416 (2020).
\bibitem{Liu2020-2} Liu, J., Wang., L. \& Shen, K.
				Dipolar spin waves in uniaxial easy-axis antiferromagnets: A natural topological nodal-line semimetal.
				\textit{Phys. Rev. Research} \textbf{2}, 023282 (2020).
\bibitem{Luttinger1946} Luttinger, J.M. \& Tisza, L.
					Theory of dipole interaction in crystals.
					\textit{Phys. Rev.} \textbf{70}, 954 (1946).
\bibitem{Mermin1966} Mermin,  N.~D. \& Wagner, H.
				Absence of ferromagnetism or antiferromagnetism in one- or two-dimensional isotropic Heisenberg models.
				\textit{Phys. Rev. Lett.} \textbf{17}, 1133 (1966).
\bibitem{Malozovsky1990} Malozovsky, Y.M. \& Rozenbaum, V.M.
					Orientational ordering in two-dimensional systems with long-range interaction.
					\textit{Physica} A \textbf{175}, 127 (1991).
\bibitem{Zimmerman1988} Zimmerman, G.O., Ibrahim A.K. \& Wu, F.Y.
					 Planar classical dipolar system on a honeycomb lattice.
					\textit{Phys. Rev. B}  \textbf{37}, 2059 (1988).
\bibitem{Rozenbaum1995} Rozenbaum, V.M.
					Ground state and vibrations of dipoles on a honeycomb lattice.
					\textit{Phys. Rev. B} \textbf{51}, 1290 (1995).
\bibitem{Maksymenko2015} Maksymenko, M., Chandra, V.R. \& Moessner, R.
					Classical dipoles on the kagome lattice.
					\textit{Phys. Rev. B} \textbf{91}, 184407 (2015).
\bibitem{Kraemer1999} Kr\"amer, K.W. \textit{et al.} 
				Noncollinear two- and three-dimensional magnetic ordering in the honeycomb lattices of ErX$_3$ (X = Cl, Br, I).
				   \textit{Phys. Rev. B} \textbf{60}, R3724 (1999).
\bibitem{Braekken1930} Braekken, H. 
				\textit{Z. Krist.} \textbf{75}, 574 (1930).
\bibitem{Bertaut1963} Bertaut, E.F. in \textit{Magnetism}, Ed. G.T. Rado Suhl and H. Suhl, vol. III, p. 150 (1963).
\bibitem{Rastelli1999} Rastelli, E.,Carbognani, A., Regina, S. \& Tassi, A.
				Order by thermal disorder in 2D planar rotator model with dipolar interactions.
				\textit{Eur. Phys. J. B} \textbf{9}, 641 (1999). 

\bibitem{Enjalran2004} Enjalran, M. \& Gingras, M.J.P. 
				Theory of paramagnetic scattering in highly frustrated magnets with long-range dipole-dipole interactions: The case of the
				 Tb$_2$Ti$_2$O$_7$ pyrochlore antiferromagnet.
				\textit{Phys. Rev. B} \textbf{70}, 174426 (2004).
\bibitem{Jensen1991} Jensen, J. \& Macintosh, A.R. \textit{Rare Earth Magnetism}, Clarendon Press, Oxford (1991).
\bibitem{Wehling2014} Wehling, T.O., Black-Schaffer, A.M. \& Balatsky, A.V. 
				Dirac materials.
				\textit{Adv. Physics} \textbf{63}, 1  (2014). 
\bibitem{Franson2016} Fransson, J., Black-Schaffer, A.M. \& Balatsky, A.V. 
				Magnon Dirac materials.
				\textit{Phys. Rev. B} \textbf{94}, 075401 (2016).
\bibitem{Boyko2018} Boyko, D., Balatsky, A.V. \&  Haraldsen, J.T. 
				Evolution of magnetic Dirac bosons in a honeycomb lattice.
				\textit{Phys. Rev. B} \textbf{97}, 014433 (2018).
\bibitem{Pershoguba2018} Pershoguba, S.S. \textit{et al.} 
				Dirac magnons in honeycomb ferromagnets.
				\textit{Phys. Rev. X} \textbf{8}, 011010 (2018).
\bibitem{Li2017} Li, K. \textit{et al.}
				Dirac and nodal line magnons in three-dimensional antiferromagnets.
				\textit{Phys. Rev. Lett.} \textbf{119}, 247202 (2017).
\bibitem{Bao2018} Bao, S. \textit{et al.}
				Discovery of coexisting Dirac and triply degenerate magnons in a three-dimensional antiferromagnet.
				\textit{Nat. Commun.} \textbf{9}, 2591 (2018). 
\bibitem{Yao2018} Yao, W. \textit{et al.}
					Topological spin excitations in a three-dimensional antiferromagnet.
					\textit{Nat. Phys.} \textbf{14}, 1011 (2018).	
\bibitem{Yuan2020} Yuan, B. \textit{et al.}
				Dirac magnons in a honeycomb lattice quantum XY magnet CoTiO$_3$.
				\textit{Phys. Rev. X} \textbf{10}, 011062 (2020). 
\bibitem{Chen2018} Chen, L. \textit{et al.}
				Topological spin excitations in honeycomb ferromagnet CrI$_3$.
				\textit{Phys. Rev. X} \textbf{8}, 041028 (2018).	
\bibitem{Delugas2021} Delugas, P. \textit{et al.}
				Magnon-phonon interactions open a gap at the Dirac point in the spin-wave spectra of CrI$_3$ 2D magnets.
				\textit{arXiv:2105.04531} (20021).
\bibitem{Hasan2010} Hasan, M.Z. \& Kane, C.L.
				Colloquium: Topological insulators.
				\textit{Rev. Mod. Phys.} \textbf{82}, 3045 (2010).
\bibitem{Xiao2007} Xiao, D., Yao, W. \& Niu, Q. 
                   Valley-contrasting physics in graphene: magnetic moment and topological transport. 
                   \textit{Phys. Rev. Lett.} \textbf{99}, 236809 (2007) . 
\bibitem{Yao2009} Yao, W., Yang, S.A. \& Niu, Q. 
                  Edge states in graphene: From gapped flat-band to gapless chiral modes.
                  \textit{Phys. Rev. Lett.} \textbf{102}, 096801 (2009).
\bibitem{Kai2018} Qian, K., Apigo, D.J., Prodan, C., Barlas, Y., Prodan, E.
                  Topology of the valley-Chern effect. 
                  \textit{Phys. Rev. B} \textbf{98}, 155138 (2018).
\bibitem{Ghader2020} Ghader, D.
			 Valley-polarized domain wall magnons in 2D ferromagnetic bilayers. 
                     \textit{Scientific Reports} \textbf{10}, 16733 (2020).
\bibitem{Zhai2020} Zhai, X. \&  Blanter, Y.M. 
                   Topological valley transport of gapped Dirac magnons in bilayer ferromagnetic insulators.
                   \textit{Phys. Rev. B} \textbf{102}, 075407 (2020).
\bibitem{Mook 2015} Mook, A., Henk, J. \& Mertig, I. 
				Topologically nontrivial magnons at an interface of two kagome ferromagnets.
					\textit{Phys. Rev. B}, \textbf{91}, 224411 (2015).

\bibitem{Pich 1995} Pich, C. \& Schwabl, F.
				Spin-wave dynamics of two-dimensional isotropic dipolar honeycomb antiferromagnets.
				\textit{JMMM} \textbf{148}, 30-31 (1995).			
\bibitem{Streubel2018} Streubel, R. \textit{et al.}
					Spatial and temporal correlations of XY macro spins.
					\textit{Nano. Lett.} \textbf{18}, 12, 7428-7434 (2018).
\bibitem{Lenk2011} Lenk, B., Ulrichs, H., Garbs, F. \& Münzenberg, M.
				 The building blocks of magnonics. 
				 \textit{Physics Reports} \textbf{507}, 107 (2011).
\end{thebibliography}
\end{document}